\documentclass[aps,prl,twocolumn,showpacs,superscriptaddress,groupedaddress]{revtex4-1}
\usepackage{natbib}
\usepackage{placeins}
\usepackage{graphicx}
\usepackage{rotating}
\usepackage{subfigure}
\usepackage{amssymb}
\usepackage{amsmath}
\usepackage{mathrsfs}
\usepackage[scientific-notation=true]{siunitx}
\usepackage{color}
\usepackage{epstopdf}
\usepackage{float}

\graphicspath{{./}} 

\newcommand{\Ca}{\textrm{Ca}}

\newcommand{\gammadot}{\dot{\gamma}}

\DeclareMathAlphabet{\mathpzc}{OT1}{pzc}{m}{it} 
\newcommand{\mrm}{\mathrm}
\newcommand{\Kpm}{K_{pm}}
\newcommand{\Kpc}{K_{pc}}
\newcommand{\Kpd}{K_{pd}}
\newcommand{\Ktm}{K_{tm}}
\newcommand{\Ktc}{K_{tc}}
\newcommand{\Ktd}{K_{td}}
\newcommand{\kpm}{\kappa_{pm}}
\newcommand{\kpc}{\kappa_{pc}}
\newcommand{\kpd}{\kappa_{pd}}
\newcommand{\ktm}{\kappa_{tm}}
\newcommand{\ktc}{\kappa_{tc}}
\newcommand{\ktd}{\kappa_{td}}

\newcommand{\kp}{\eta_p}

\newcommand{\conf}{\mrm{C}}
\newcommand{\fancyD}{\mathpzc{D}}
\newcommand{\phip}{\phi_{p}}
\newcommand{\phit}{\phi_{t}}
\newcommand{\phipc}{\phi_{pc}}
\newcommand{\phitc}{\phi_{tc}}
\newcommand{\marg}{\textrm{M}}
\newcommand{\rratio}{\textrm{F}}
\newcommand{\sratio}{\textrm{S}}

\makeatletter

\newcommand{\numtoRoman}[1]{\expandafter\@slowromancap\romannumeral #1@}
\makeatother

\begin{document}

\title{Margination regimes and drainage transition in confined multicomponent suspensions}
\author{Rafael G.~{Henr\'iquez Rivera}}
\author{Kushal Sinha}
\author{Michael D.~Graham}\email{Corresponding author. E-mail: mdgraham@wisc.edu}
\affiliation{
Department of Chemical and Biological Engineering\\
University of Wisconsin-Madison, Madison, WI 53706-1691
}
\date{\today}

\begin{abstract}
A mechanistic theory is developed to describe segregation in confined multicomponent suspensions such as blood. It incorporates the two key phenomena arising in these systems at low Reynolds number: hydrodynamic pair collisions and wall-induced migration. In simple shear flow, several regimes of segregation arise, depending on the value of a ``margination parameter'' $\marg$. Most importantly, there is a critical value of $\marg$ below which a sharp ``drainage transition'' occurs:  one component is completely depleted from the bulk flow to the vicinity of the walls. Direct simulations also exhibit this transition as the size or flexibility ratio of the components changes. 
\end{abstract}

\maketitle

\paragraph{Introduction.}Flow-induced segregation is ubiquitous in multicomponent suspensions and granular materials, including systems as disparate as hard macroscopic particles in air \cite{Mobius:2001vh}, polydisperse droplet suspensions \cite{Makino:2013fz},  foams \cite{Mohammadigoushki:2013iu}, and blood. During blood flow,  the focus of the present work, both the leukocytes and platelets segregate near the vessel walls, a phenomenon known as margination, while the red blood cells (RBCs) tend to be depleted in the near-wall region, forming a so-called cell-free  or depletion layer \cite{Sutera:1993un,*tangelder85,*lipowsky89,*popel05,*Kumar:2012ga,*Grandchamp:2013jq}.  Engineering the margination process has been proposed for microfluidic cell separations in blood (\emph{e.g.}~\cite{WeiHou:2012is}) as well as for enhanced drug delivery to the vasculature \cite{Namdee:2013fc,*Thompson:2013dm}. 

Direct simulations of flowing multicomponent suspensions  -- models of blood -- can capture margination phenomena \cite{Freund:2007kx,Crowl:2011cf,Zhao:2011do,Kumar:2011dd,Fedosov:2012dy,Reasor:2012ey,Zhao:2012gg,Fedosov:2013ie,Vahidkhah:2014hy,Kumar:2013tu}, but  developing a fundamental understanding of underlying mechanisms and parameter-dependence from simulations is difficult. It is thus important to have a simple yet mechanistic mathematical model, ideally one with closed form solutions that reveal parameter-dependence, that can distill out the essential phenomena that drive segregation and capture the key effects and transitions.  We present such a model here.

\begin{figure}[ht]
  \centering
        \includegraphics[width=3.375in]{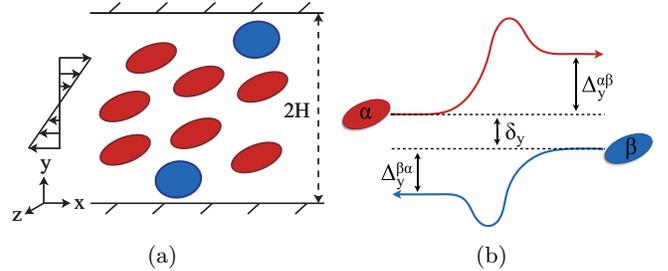}
        
        \caption{(a) Suspension of particles in a slit under simple shear flow. (b) Pair collision trajectories of particles of species $\alpha$ and $\beta$ under simple shear flow.}
        \label{fig:schematic}
\end{figure}

\paragraph{Theory.}
We consider a dilute suspension containing $N_{s}$ types of deformable particles with total volume fraction $\phi$ undergoing flow in a slit bounded by no-slip walls at $y=0$ and $y=2H$ and unbounded in $x$ and $z$. Quantities referring to a specific component $\alpha$ in the mixture will have subscript $\alpha$: for example $n_{\alpha}$ is the number density of component $\alpha$.  We consider here only simple shear (plane Couette) flow and, consistent with the diluteness assumption, take the shear rate $\gammadot$ to be independent of the local number densities and thus independent of position. In a dilute suspension of particles, where $\phi \ll 1$, the particle-particle interactions can be treated as a sequence of uncorrelated pair collisions \cite{cunha96,Li:2000wv,zurita12}. 
For the moment, we neglect molecular diffusion of the particles. This issue is further addressed below.
Since the particles are deformable, they migrate away from the wall during flow with velocity $v_{\alpha m}(y)$ \cite{Smart:1991vp,Ma:2005dw}.  The evolution of the particle number density distributions can be idealized by a kinetic master equation that captures  the migration and collision effects (\cite{zurita12,Kumar:2012ie,Narsimhan:2013jk,Kumar:2013tu}).  Assuming uniform particle distributions in $x$ and $z$,  this equation is
\begin{multline}\label{eq:meq}
\frac{\partial n_{\alpha}(y,t)}{\partial t}  =  - \frac{\partial}{\partial y} \big(v_{\alpha m}(y) n_{\alpha}(y,t)\big)\\
+ \,  \sum_{\beta=1}^{N_s}   \bigg(  \int_{-(2H-y)}^{y} \int_{-\infty}^{\infty}   \bigg\{ n_{\alpha}(y-\Delta_y^{\alpha\beta},z-\Delta_z^{\alpha\beta},t) \\
\times n_\beta(y-\Delta_y^{\alpha\beta}-\delta_y,z-\Delta_z^{\alpha\beta}-\delta_z,t) \\
- \,  n_\alpha(y,z,t) n_\beta(y-\delta_y,z-\delta_z,t) \bigg\} \, \dot{\gamma} \, |\delta_{y}| \, d\delta_z d\delta_y \bigg).
\end{multline}
Here $\delta_y$ and $\delta_z$ are the pre-collision pair offsets in the $y$ and $z$ directions
and $\Delta_y^{\alpha\beta}(\delta_y,\delta_z)$ and $\Delta_z^{\alpha\beta}(\delta_y,\delta_z)$ are
the cross-stream and cross-vorticity direction displacements of a particle of type $\alpha$ after collision with a
particle of type $\beta$. 
See Fig.~\ref{fig:schematic} for a schematic. 
The term $\dot{\gamma}|\delta_y|$ in the integrand accounts for the relative velocity of
approach of two colliding particles. 

We now construct an approximation to this model that is valid in the limit $(\Delta_y^{\alpha\beta},\Delta_z^{\alpha\beta})\rightarrow 0$. 
Taylor-expanding the first term in the curly brackets in Eq.~\ref{eq:meq}  about
$\Delta_y^{\alpha\beta}=~\Delta_z^{\alpha\beta}=0$, neglecting terms involving $(\Delta^{\alpha\beta})^3$ and smaller, and applying the condition that $n_{\alpha}$ is independent of $z$
yields a set of nonlocal drift-diffusion equations: 
\begin{equation}\label{eq:fpe}
\frac{\partial n_{\alpha}}{\partial t} = - \frac{\partial}{\partial y} \left( \left(v_{\alpha m}+v_{\alpha c}\right)n_{\alpha} - \frac{\partial}{\partial y}(D_{\alpha}n_{\alpha}) \right).
\end{equation}
Here $v_{\alpha c}$ is the collisional drift velocity of component $\alpha$, while $D_{\alpha}$ is its short time
self-diffusivity. In the important special case of a binary suspension composed of a ``primary'' component ($\alpha=$`$p$') and a ``trace'' component ($\alpha=$`$t$') such that $n_{p}\gg n_{t}$, only the primary component contributes to these quantities: 
\begin{equation}
v_{\alpha c} (y)   =   \int_{-r_{cut}}^{r_{cut}} n_{p}(y-\delta_y)   \widehat{\Delta_y^{\alpha p}}(\delta_y)   \dot{\gamma} |\delta_y| ~ d\delta_y ,
\label{eq:vintegrals}
\end{equation}
\begin{equation}
D_{\alpha}(y) = \frac{1}{2}  \int_{-r_{cut}}^{r_{cut}} n_{p}(y-\delta_y)  \widehat{\left(\Delta_y^{\alpha p}\right)^{2}}(\delta_y)  \dot{\gamma} |\delta_y| ~ d\delta_y.
\label{eq:dintegrals}
\end{equation}
Here $r_{cut}$ is the radius beyond which particle-particle interaction is assumed to be negligible and  
\begin{equation}\label{eq:z_integral}
 \widehat{\Delta_y^{\alpha p}}(\delta_y) = \int_{-r_{cut}}^{r_{cut}} \Delta_y^{\alpha p}(\delta_y,\delta_z) d\delta_z,
\end{equation}
\begin{equation}\label{eq:z_integral2}
 \widehat{\left(\Delta_y^{\alpha p}\right)^{2}}(\delta_y) = \int_{-r_{cut}}^{r_{cut}} \{\Delta_y^{\alpha p}(\delta_y,\delta_z) \}^{2}d\delta_z.
\end{equation}
The condition $n_{p}\gg n_{t}$ is valid for blood, where RBCs outnumber platelets and white blood cells by one and three orders of magnitude respectively \cite{Boal:2012wq}.

Finally, a further simplification allows substantial additional insight. 
We make local approximations to the integrals, Eqs.~\ref{eq:vintegrals} and \ref{eq:dintegrals}, based on the argument that $\Delta_y^{tp}$ and $\Delta_y^{pp}$ are vanishingly small for large $|\delta_y|$ by Taylor-expanding $n_{p}$ around $\delta_y=0$, noting that $\Delta_y(\delta_y,\delta_{z})$ is odd in $\delta_{y}$ and keeping only the leading terms. Now the collisional drift velocities and diffusivities become
\begin{align}
v_{\alpha c}&=-K_{\alpha c}\frac{\partial \gammadot n_{p}}{\partial y}, \quad D_{\alpha}=K_{\alpha d}\gammadot n_{p},
\end{align}
where
\begin{align}
K_{\alpha c}&=2\int_{0}^{r_{cut}} \widehat{\Delta_y^{\alpha p}}(\delta_y)\delta_y |\delta_y| \;d\delta_y,  \label{eq:Kpc}\\
K_{\alpha d}&=\int_{0}^{r_{cut}} \widehat{\left(\Delta_y^{\alpha p}\right)^{2}}(\delta_y)|\delta_y| \;d\delta_y. \label{eq:Kpd}
\end{align}
The convergence of these integrals deserves mention. In the far field each particle appears as a force dipole, so in an unbounded domain the collisional displacements $\Delta_{y}^{\alpha p}$ would decay as $\delta_{y}^{-2}$. Thus convergence of Eq.~\ref{eq:Kpd} is unproblematic irrespective of $r_{cut}$. For convergence of Eq.~\ref{eq:Kpc}, $r_{cut}$ must be bounded. An explicit bound is the slit width $2H$. Furthermore, at any finite concentration the spacing between particles scales as $n_{p}^{-1/3}$. A given particle will effectively only collide with other particles within this range, while particles outside this range would be more strongly affected by their nearer neighbors.   

To describe the wall-induced hydrodynamic migration velocity, we superpose the point-force-dipole approximations corresponding to each of the two walls \cite{Pranay:2012ku}:
\begin{align} \label{eq:dipole}
v_{\alpha m}&=K_{\alpha m} \left( \frac{1}{y^2} - \frac{1}{(2H-y)^2} \right).
\end{align}
The parameter $K_{\alpha m}$ depends linearly on the $yy$-component of the stresslet generated by the deformable particle \cite{Smart:1991vp,Ma:2005dw}. This scales as $a_{\alpha}^{4}$, where $a_{\alpha}$ is the particle radius of species $\alpha$, and as $\gammadot^{2}$ at low $\gammadot$ with this dependency becoming weaker as $\gammadot$ increases \cite{Kumar:2013tu,Pranay:2012ku}.

With these further idealizations, Eq.~\ref{eq:fpe} becomes a pair of partial differential equations, which we present here in nondimensional form:
\begin{align} \label{eq:fpe-pnd} 
    \frac{\partial \phi_p}{\partial t} &= -\frac{\partial}{\partial y}\left[ \kpm \left( \frac{1}{y^2} - \frac{1}{(2\mathrm{C}-y)^2} \right) \phi_p \right. \\ \nonumber
    &\left. - \, \kpc \frac{\partial \phi_p}{\partial y} \phi_p - \kpd \frac{\partial \phi_p^2}{\partial y} \right], \\ \label{eq:fpe-tnd}
    \frac{\partial \phi_t}{\partial t} &= -\frac{\partial}{\partial y} \left[ \ktm \left( \frac{1}{y^2} - \frac{1}{(2\mathrm{C}-y)^2} \right) \phi_t \right. \\ \nonumber
    &\left. - \, \ktc \frac{\partial \phi_p}{\partial y} \phi_t - \ktd \frac{\partial (\phi_p \phi_t)}{\partial y} \right].
\end{align}
Here $\phi_p=n_pV_p$ and $\phit = n_tV_t$  are the volume fractions of the primary and trace components, 
where $V_{\alpha}$ is the volume per particle of component $\alpha$, 
$\mrm{C}=H/a_p$ is the confinement ratio,  $\kpm = \frac{\Kpm}{\dot{\gamma}a_p^3}$, $\kpc = \frac{\Kpc}{V_pa_p^2}$, $\kpd = \frac{\Kpd}{V_pa_p^2}$, $\ktm = \frac{\Ktm}{\dot{\gamma}a_p^3}$, $\ktc = \frac{\Ktc}{V_pa_p^2}$, and $\ktd = \frac{\Ktd}{V_pa_p^2}$. Time $t$ is nondimensionalized with $\dot{\gamma}^{-1}$ and $y$ with $a_p$. For simplicity, we keep the symbols $t$ and $y$ for their nondimensionalized forms. 
For a single-component suspension of rigid particles ($N_{s}=1,K_{\alpha m}=0$)  a model of similar form was proposed by \cite{PHILLIPS:1992vy} based on phenomenological arguments first proposed by \cite{LEIGHTON:1987uo}.

\paragraph{Results.}

\begin{figure}[hb]
  \centering
        \includegraphics[width=3.375in]{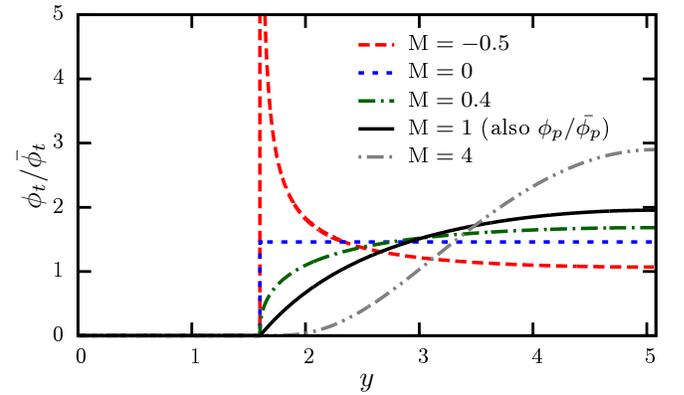}
        \caption{Steady state volume fraction profiles of $\phi_{p}/\bar{\phi}_{p}$ (black solid line) and $\phi_{t}/\bar{\phi}_{t}$ for various values of $\marg$. (The curves coincide when $\marg=1$.) Here $\bar{\phi}_{p}=0.12$, $\phipc = 0.23$, $\conf = 5.08$, $\kpm = 0.11$, $\kpc = 0.02$, and $\kpd = 0.07$, resulting in $l_d=1.6$ (extracted from simulation results in \cite{Kumar:2013tu}). For simplicity, $\ktd = \kpd$ and $\ktc = \kpc$. We vary $\marg$  by changing $\ktm$.}
        \label{fig:phit_an}
\end{figure}

An important feature of Eqs.~\ref{eq:fpe-pnd} and \ref{eq:fpe-tnd} is that steady state solutions with no-flux boundary conditions at the wall ($y=0$) and centerline ($y=\mrm{C}$), can be found analytically. For $\phi_{p}$ we find
\begin{equation} \label{eq:phi_p}
	\phip=\left\{\begin{matrix} 0, &  y<l_d \\
\phipc\left(1-\frac{2\kp}{\mrm{C}\phipc}\frac{(\mrm{C}-y)^2}{y(2\mrm{C}-y)}\right), &y>l_d \end{matrix}\right. ,
\end{equation}
where $\phipc$ is the volume fraction of the primary component at the centerline, $\kp = \frac{\kpm}{\kpc+2\kpd}$, and $l_d$ is the nondimensional cell-free layer thickness:
\begin{equation}\label{eq:l_d}
	l_d = \mrm{C} \left( 1 - \sqrt{\frac{\mrm{C}\phipc}{2\kp + \mrm{C}\phipc}} \right) .
\end{equation}
The black solid line in Fig.~\ref{fig:phit_an} shows $\phip$ normalized by its mean volume fraction $\bar{\phip}$.  In the unconfined limit $\conf\rightarrow\infty$, $l_{d}\rightarrow\kp/\phipc$, confirming the $\phi^{-1}$ dependence found earlier in scaling analyses \cite{Hudson:2003df,Pranay:2012ku,Narsimhan:2013jk}. More generally, Eq.~\ref{eq:l_d} analytically captures the dependence of the cell-free layer thickness on the volume fraction, degree of confinement and particle properties.

For the trace component, the steady state solution is 
\begin{align}\label{eq:phi_t}
	\phit=\left\{\begin{matrix} 0, &  y<l_d \\
 \phitc\left( \frac{\phip(y)}{\phipc}\right)^\marg, &y>l_d \end{matrix}\right. ,
\end{align}
where $\phip(y)$ is the steady state solution found above, $\phitc$ is the centerline volume fraction of the trace component and 
\begin{equation}\label{eq:Gamma}
	\marg = \frac{\kpc+2\kpd}{\ktd}\left( \frac{\ktm}{\kpm} - \frac{\ktc+\ktd}{\kpc+2\kpd}\right).
\end{equation}
Remarkably, this single quantity, which we call the \emph{margination parameter}, determines the qualitative nature of the concentration profile.

The sign of $\marg$ is determined by the competition between the ratio of the migration velocities of the two components, $\frac{\ktm}{\kpm}$, and the ratio of the collisional terms, $\frac{\ktc+\ktd}{\kpc+2\kpd}$.
Depending on  $\marg$, several distinct regimes of behavior can be identified:\\
(1) $\marg > 1$: the trace component is displaced further from the wall than the primary component: it \emph{demarginates}. \\
(2) $0 < \marg < 1$: the relative concentration of the trace component is higher near the wall than the primary component but does not display a peak: it \emph{weakly marginates}. \\
(3) $-1 < \marg < 0$: the trace component displays a peak at $y = l_d$, corresponding to an integrably singular concentration profile: it \emph{moderately marginates}. \\
(4) $\marg \leq -1$: here Eq.~\ref{eq:phi_t}  displays a nonintegrable singularity at $y = l_d$. This steady state is physically unrealizable as it corresponds to an infinite amount of material in a finite region. In this regime collisional transport overwhelms migration, and the trace component accumulates indefinitely at $y=l_{d}$, indicating \emph{strong margination}.

\begin{figure}[ht]
  \centering
       \includegraphics[width=3.375in]{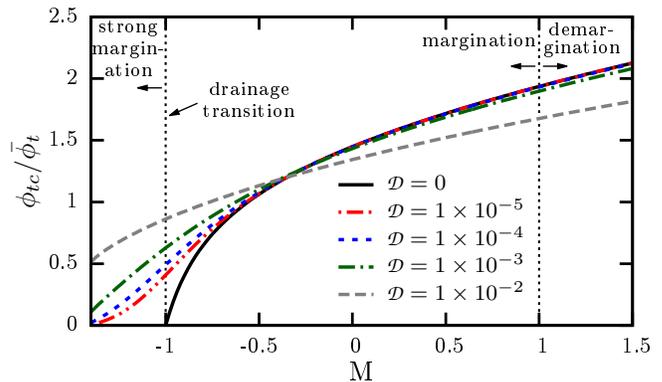}
        \caption{Centerline volume fraction of the trace component $\phi_{tc}$ scaled with the average trace concentration $\bar{\phit}$ \emph{vs.}~$\marg$ for varying $\mathpzc{D}$. Other parameters are the same as in Fig.~\ref{fig:phit_an}.}
        \label{fig:phitc_an}
\end{figure}

The black solid line in Fig.~\ref{fig:phitc_an} shows the ratio between the centerline concentration $\phitc$ and the average concentration $\bar{\phit}$ \emph{vs.}~$\marg$. This falls sharply to zero at $\marg = -1$; we call this phenomenon the \emph{drainage transition}, since for $\marg\leq -1$ all the trace component is completely drained from the bulk. If the trace component does not migrate (as in the case of rigid particles), then $\ktm=0$ and  $\marg=-(1+\kappa_{tc}/\kappa_{td})$, which is \emph{always} less than $-1$. (This case is degenerate in the absence of Brownian diffusion, because at steady state $\phi_{t}$ can take on arbitrary values when $y<l_{d}$.)

For particles the size of blood cells ($> 1 \mu m$) at shear rates characteristic of the microcirculation ($10^{2}-10^{3}s^{-1}$), Brownian diffusion is unimportant. For smaller particles, however, such as might be used for drug delivery, this may no longer be true. We  consider the impact of Brownian diffusion on trace component transport by adding an appropriately nondimensionalized diffusion term $\fancyD\;\partial^{2}\phi_{t}/\partial y^{2}$ to Eq.~\ref{eq:fpe-tnd}. Here $\mathpzc{D} = D_{B}/a_{p}^{2}\gammadot$, where $D_{B}$ is the Brownian diffusivity of the trace component. 
Using typical values for blood ($a_p = \num{0.0000039}$~m, $\gammadot = 500$~s$^{-1}$) and the Stokes-Einstein relation, varying $\fancyD$ from $10^{-5}$ to $10^{-2}$ corresponds to varying $a_t$ from $\sim 10^{-6}\textrm{ m}$ to $\sim 10^{-9}\textrm{m}$. 

The steady solution for the trace component with molecular diffusion is
\begin{equation}\label{eq:phi_tpe}
	\phit=\left\{\begin{matrix} \phitc\left(1-\frac{2\kp\ktd}{\mrm{C}(\mathpzc{D}+\phipc\ktd)}\frac{(\mrm{C}-l_d)^2}{l_d(2\mrm{C}-l_d)}\right)^\marg &  \\ 
	\times\exp\left( -\frac{2\ktm\mrm{C}}{\mathpzc{D}}\left( \frac{1}{y(2\mrm{C}-y)} - \frac{1}{l_d(2\mrm{C}-l_d)} \right) \right), & y<l_d \\ 
	\phitc\left(1-\frac{2\kp\ktd}{\mrm{C}(\mathpzc{D}+\phipc\ktd)}\frac{(\mrm{C}-y)^2}{y(2\mrm{C}-y)}\right)^\marg, &y>l_d. \end{matrix}\right.
\end{equation}
Molecular diffusion results in a spreading of $\phit$ to include the region $y<l_{d}$ and also renders the steady solution for $\marg \leq -1$ integrable.  
It also smears out the drainage transition as shown in Fig.~\ref{fig:phitc_an}.
\begin{figure}[ht]
  \centering
        \includegraphics[width=3.375in]{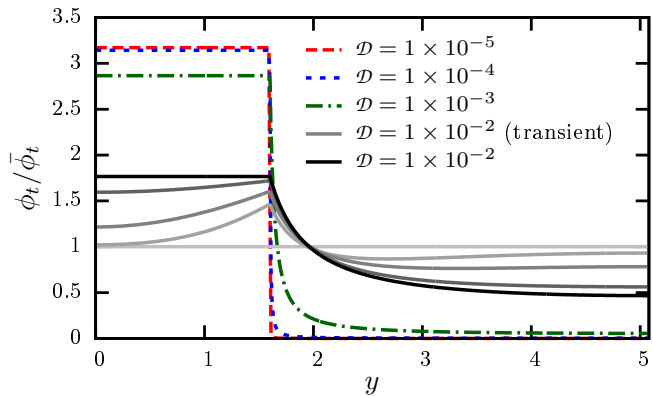}
        \caption{Steady state volume fraction profiles $\phi_{t}(y)/\bar{\phi}_{t}$ for $\ktm = 0$ and various $\mathpzc{D}$. Other parameters are the same as in Fig. 2. Also, the solid lines show the transient evolution for $\fancyD = \num{0.01}$: $t=0$ (lightest gray), 30, 90, 270, and at steady state (black).}
        \label{fig:phit_an_ktm0}
\end{figure}

Now consider the rigid trace particle case $\ktm = 0$. 
  Fig.~\ref{fig:phit_an_ktm0} shows how the steady state profile of $\phit$ varies with $\fancyD$:
  Margination is weakened by diffusion. 
  Fig.~\ref{fig:phit_an_ktm0} also shows the transient evolution of $\phit$ for $\fancyD = \num{0.01}$ from a uniform initial condition as determined from a numerical simulation using a conservative finite volume method.  

\begin{figure}[hb]
  \centering
	\includegraphics[width=3.375in]{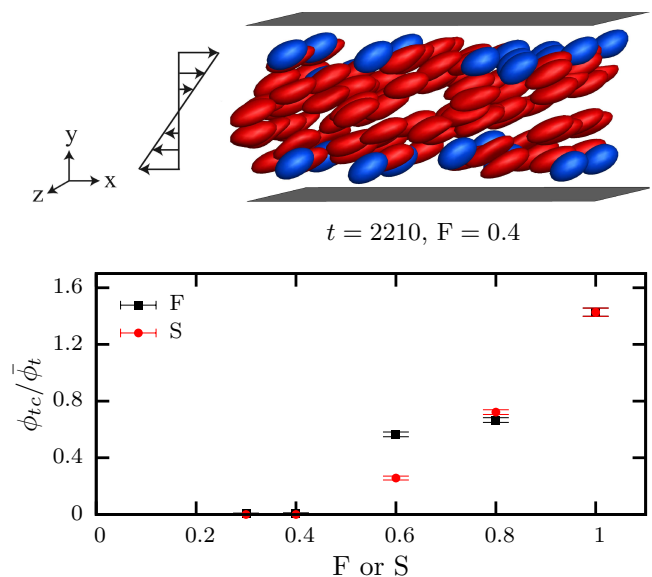}
	\caption{Time-averaged volume fraction of the trace components at the centerline as a function of flexibility ratio $\rratio$ or size ratio $\sratio$ from direct hydrodynamic simulations, where $\conf=5.08$ and $l_{d}=1.6$. In case (a)  $\bar{\phi}=0.12$ and in case (b) $\bar{\phi}~=~\{0.096, 0.098, 0.101, 0.108, 0.12\}$ for $\sratio~=~\{0.2, 0.4, 0.6, 0.8, 1.0\}$.}
	\label{fig:BIM}
\end{figure}
  
\paragraph{Comparison with direct simulations.}
To evaluate the prediction of a drainage transition, we performed direct simulations of binary suspensions of fluid-filled non-Brownian elastic capsules at low Reynolds number using a boundary integral method (cf.~\cite{Kumar:2011dd,Kumar:2012ev,Kumar:2013tu}).  Two cases were considered: segregation by (a) deformability  and (b) size. 
The particles are all spherical at rest. Particle deformability is characterized by the capillary number $\textrm{Ca}_{\alpha} = \mu \dot{\gamma} a_\alpha/G_{\alpha}$,  where $\mu$ is the fluid viscosity and $G_{\alpha}$ is the membrane shear modulus of component $\alpha$. In case (a) the primary  component comprises 80\% of the particles and has $\textrm{Ca}_{p}=0.5$; the trace component is stiffer, and we define a flexibility ratio $\rratio=G_{p}/G_{t}=\Ca_{t}/\Ca_{p}$. The primary component in case (b) is the same as in case (a), but now the trace component is smaller as defined by the size ratio $\sratio=a_{t}/a_{p}$. In this case $\Ca_{t}=\Ca_{p}$. 

Fig.~\ref{fig:BIM} shows the steady-state value of $\phi_{tc}/\bar{\phi}_{t}$  as $\rratio$ or $\sratio$ changes.  It is very similar to Fig.~\ref{fig:phitc_an}, clearly indicating that the drainage transition predicted by theory is found in the simulations. Coincidentally, the transition is in the same range $0.4-0.6$ for both $\sratio$ and $\rratio$ under the conditions chosen.  Considering case (b) first, the migration parameter $\kappa_{tm}$ scales as $\sratio^{3}$ at constant $\Ca_{t}$, so as $\sratio$ decreases so does $\marg$; recall that $\marg<-1$ for $\kappa_{tm}=0$. With regard to case (a), $\ktm$ also decreases with decreasing $\rratio$, and additionally the collisional displacements and thus $\ktc$ and $\ktd$ increase \cite{Kumar:2011dd}. Therefore, decreasing $\rratio$ also corresponds to decreasing $\marg$, resulting in a drainage transition.
\begin{figure}[hb]
  \centering
        \includegraphics[width=3.375in]{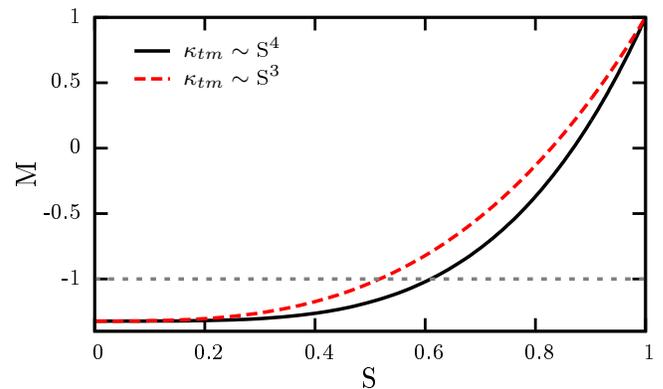}
        \caption{$\marg$ \emph{vs.} $\sratio$ for $\ktm\sim\sratio^4$ (black solid line) and $\ktm\sim\sratio^3$ (red dashed line); the latter corresponds to the simulations shown in Fig.~\ref{fig:BIM}.
          Other parameters are the same as in Fig. 2.}
        \label{fig:M_vs_S}
\end{figure}

Returning to the theory, using the values in the caption of Fig.~\ref{fig:phit_an} and Eq.~\ref{eq:Gamma}, we can determine the value of $\sratio$ corresponding to the drainage transition by finding $\marg$ as $\sratio$ is varied. This result is shown in Fig.~\ref{fig:M_vs_S}, where we use $\ktm\sim\sratio^4$ (black solid line) and $\ktm\sim\sratio^3$ (red dashed line) to represent the cases of varying $a_t$ while keeping $G_t$ and $\Ca_t$ constant, respectively. The values of $\sratio$ corresponding to the drainage transition are 0.61 and 0.52, respectively. The latter case corresponds to case (b) above, and we see that the theory result agrees well with the direct simulation result in Fig.~\ref{fig:BIM}.

For reference to blood, the values of $\sratio$ for leukocytes and platelets with respect to RBCs  are $0.9-2$ \cite{schmid1980morphometry} and $0.25$ \cite{paulus1975platelet} respectively, while $\rratio$ is of the order $10^{-2}$ \cite{schmid1981passive} and $10^{-4}$ \cite{lam2011mechanics} respectively. Thus, case (a) here is more closely related to the RBC-leukocyte segregation, where the size ratio is close to unity, and case (b) more nearly represents the RBC-platelet case, where the sizes are very different. From the present results it appears that both leukocytes and platelets would satisfy the conditions for drainage in simple shear.  

\paragraph{Conclusions.}
A mechanistic theory has been developed for the first time to describe flow-induced segregation phenomena in multicomponent suspensions such as blood. Experimental and computational observations of margination and demargination are captured qualitatively in simple closed form solutions. Several different margination regimes arise and a sharp {drainage transition} is identified beyond which the trace component of the suspension partitions completely to the edge of the cell-free layer. Direct simulations corroborate this prediction.

The framework presented here can be extended in many directions. For example, pressure-driven flow, which is common in microfluidic and circulatory applications, can be studied. The model can be expanded to include many other phenomena, including red blood cell aggregation and platelet adhesion. Most importantly, the mechanistic nature of the theory leads to substantial and systematic insight into the origins of margination; this will complement detailed simulations and experiments in guiding development of technologies involving blood and other multicomponent suspensions at small scales. 

This material is based upon work supported by the National Science Foundation under Grants No.~CBET-1132579 and No.~CBET-1436082, the National Science Foundation Graduate Research Fellowship Program under Grant No.~DGE-1256259 granted to RH, and a BP graduate fellowship granted to KS.

%


\begin{thebibliography}{34}%
\makeatletter
\providecommand \@ifxundefined [1]{%
 \@ifx{#1\undefined}
}%
\providecommand \@ifnum [1]{%
 \ifnum #1\expandafter \@firstoftwo
 \else \expandafter \@secondoftwo
 \fi
}%
\providecommand \@ifx [1]{%
 \ifx #1\expandafter \@firstoftwo
 \else \expandafter \@secondoftwo
 \fi
}%
\providecommand \natexlab [1]{#1}%
\providecommand \enquote  [1]{``#1''}%
\providecommand \bibnamefont  [1]{#1}%
\providecommand \bibfnamefont [1]{#1}%
\providecommand \citenamefont [1]{#1}%
\providecommand \href@noop [0]{\@secondoftwo}%
\providecommand \href [0]{\begingroup \@sanitize@url \@href}%
\providecommand \@href[1]{\@@startlink{#1}\@@href}%
\providecommand \@@href[1]{\endgroup#1\@@endlink}%
\providecommand \@sanitize@url [0]{\catcode `\\12\catcode `\$12\catcode
  `\&12\catcode `\#12\catcode `\^12\catcode `\_12\catcode `\%12\relax}%
\providecommand \@@startlink[1]{}%
\providecommand \@@endlink[0]{}%
\providecommand \url  [0]{\begingroup\@sanitize@url \@url }%
\providecommand \@url [1]{\endgroup\@href {#1}{\urlprefix }}%
\providecommand \urlprefix  [0]{URL }%
\providecommand \Eprint [0]{\href }%
\providecommand \doibase [0]{http://dx.doi.org/}%
\providecommand \selectlanguage [0]{\@gobble}%
\providecommand \bibinfo  [0]{\@secondoftwo}%
\providecommand \bibfield  [0]{\@secondoftwo}%
\providecommand \translation [1]{[#1]}%
\providecommand \BibitemOpen [0]{}%
\providecommand \bibitemStop [0]{}%
\providecommand \bibitemNoStop [0]{.\EOS\space}%
\providecommand \EOS [0]{\spacefactor3000\relax}%
\providecommand \BibitemShut  [1]{\csname bibitem#1\endcsname}%
\let\auto@bib@innerbib\@empty
\bibitem [{\citenamefont {Mobius}\ \emph {et~al.}(2001)\citenamefont {Mobius},
  \citenamefont {Lauderdale}, \citenamefont {Nagel},\ and\ \citenamefont
  {Jaeger}}]{Mobius:2001vh}%
  \BibitemOpen
  \bibfield  {author} {\bibinfo {author} {\bibfnamefont {M.~E.}\ \bibnamefont
  {Mobius}}, \bibinfo {author} {\bibfnamefont {B.~E.}\ \bibnamefont
  {Lauderdale}}, \bibinfo {author} {\bibfnamefont {S.~R.}\ \bibnamefont
  {Nagel}}, \ and\ \bibinfo {author} {\bibfnamefont {H.~M.}\ \bibnamefont
  {Jaeger}},\ }\href@noop {} {\bibfield  {journal} {\bibinfo  {journal}
  {Nature}\ }\textbf {\bibinfo {volume} {414}},\ \bibinfo {pages} {270}
  (\bibinfo {year} {2001})}\BibitemShut {NoStop}%
\bibitem [{\citenamefont {Makino}\ and\ \citenamefont
  {Sugihara-Seki}(2013)}]{Makino:2013fz}%
  \BibitemOpen
  \bibfield  {author} {\bibinfo {author} {\bibfnamefont {M.}~\bibnamefont
  {Makino}}\ and\ \bibinfo {author} {\bibfnamefont {M.}~\bibnamefont
  {Sugihara-Seki}},\ }\href@noop {} {\bibfield  {journal} {\bibinfo  {journal}
  {Biorheology}\ }\textbf {\bibinfo {volume} {50}},\ \bibinfo {pages} {149}
  (\bibinfo {year} {2013})}\BibitemShut {NoStop}%
\bibitem [{\citenamefont {Mohammadigoushki}\ and\ \citenamefont
  {Feng}(2013)}]{Mohammadigoushki:2013iu}%
  \BibitemOpen
  \bibfield  {author} {\bibinfo {author} {\bibfnamefont {H.}~\bibnamefont
  {Mohammadigoushki}}\ and\ \bibinfo {author} {\bibfnamefont {J.~J.}\
  \bibnamefont {Feng}},\ }\href@noop {} {\bibfield  {journal} {\bibinfo
  {journal} {Langmuir}\ }\textbf {\bibinfo {volume} {29}},\ \bibinfo {pages}
  {1370} (\bibinfo {year} {2013})}\BibitemShut {NoStop}%
\bibitem [{\citenamefont {Sutera}\ and\ \citenamefont
  {Skalak}(1993)}]{Sutera:1993un}%
  \BibitemOpen
  \bibfield  {author} {\bibinfo {author} {\bibfnamefont {S.~P.}\ \bibnamefont
  {Sutera}}\ and\ \bibinfo {author} {\bibfnamefont {R.}~\bibnamefont
  {Skalak}},\ }\href@noop {} {\bibfield  {journal} {\bibinfo  {journal} {Annu.
  Rev. Fluid Mech.}\ }\textbf {\bibinfo {volume} {25}},\ \bibinfo {pages} {1}
  (\bibinfo {year} {1993})}\BibitemShut {NoStop}%
\bibitem [{\citenamefont {Tangelder}\ \emph {et~al.}(1985)\citenamefont
  {Tangelder}, \citenamefont {Teirlinck}, \citenamefont {Slaaf},\ and\
  \citenamefont {Reneman}}]{tangelder85}%
  \BibitemOpen
  \bibfield  {author} {\bibinfo {author} {\bibfnamefont {G.~J.}\ \bibnamefont
  {Tangelder}}, \bibinfo {author} {\bibfnamefont {H.~C.}\ \bibnamefont
  {Teirlinck}}, \bibinfo {author} {\bibfnamefont {D.~W.}\ \bibnamefont
  {Slaaf}}, \ and\ \bibinfo {author} {\bibfnamefont {R.~S.}\ \bibnamefont
  {Reneman}},\ }\href@noop {} {\bibfield  {journal} {\bibinfo  {journal} {Am.
  J. Physiol.-Heart C}\ }\textbf {\bibinfo {volume} {248}},\ \bibinfo {pages}
  {H318} (\bibinfo {year} {1985})}\BibitemShut {NoStop}%
\bibitem [{\citenamefont {Firrell}\ and\ \citenamefont
  {Lipowsky}(1989)}]{lipowsky89}%
  \BibitemOpen
  \bibfield  {author} {\bibinfo {author} {\bibfnamefont {J.~C.}\ \bibnamefont
  {Firrell}}\ and\ \bibinfo {author} {\bibfnamefont {H.~H.}\ \bibnamefont
  {Lipowsky}},\ }\href@noop {} {\bibfield  {journal} {\bibinfo  {journal} {Am.
  J. Physiol.-Heart C.}\ }\textbf {\bibinfo {volume} {256}},\ \bibinfo {pages}
  {H1667} (\bibinfo {year} {1989})}\BibitemShut {NoStop}%
\bibitem [{\citenamefont {Popel}\ and\ \citenamefont
  {Johnson}(2005)}]{popel05}%
  \BibitemOpen
  \bibfield  {author} {\bibinfo {author} {\bibfnamefont {A.~S.}\ \bibnamefont
  {Popel}}\ and\ \bibinfo {author} {\bibfnamefont {P.~C.}\ \bibnamefont
  {Johnson}},\ }\href@noop {} {\bibfield  {journal} {\bibinfo  {journal} {Annu.
  Rev. Fluid Mech.}\ }\textbf {\bibinfo {volume} {37}},\ \bibinfo {pages} {43}
  (\bibinfo {year} {2005})}\BibitemShut {NoStop}%
\bibitem [{\citenamefont {Kumar}\ and\ \citenamefont
  {Graham}(2012{\natexlab{a}})}]{Kumar:2012ga}%
  \BibitemOpen
  \bibfield  {author} {\bibinfo {author} {\bibfnamefont {A.}~\bibnamefont
  {Kumar}}\ and\ \bibinfo {author} {\bibfnamefont {M.~D.}\ \bibnamefont
  {Graham}},\ }\href@noop {} {\bibfield  {journal} {\bibinfo  {journal} {Soft
  Matter}\ }\textbf {\bibinfo {volume} {8}},\ \bibinfo {pages} {10536}
  (\bibinfo {year} {2012}{\natexlab{a}})}\BibitemShut {NoStop}%
\bibitem [{\citenamefont {Grandchamp}\ \emph {et~al.}(2013)\citenamefont
  {Grandchamp}, \citenamefont {Coupier}, \citenamefont {Srivastav},
  \citenamefont {Minetti},\ and\ \citenamefont
  {Podgorski}}]{Grandchamp:2013jq}%
  \BibitemOpen
  \bibfield  {author} {\bibinfo {author} {\bibfnamefont {X.}~\bibnamefont
  {Grandchamp}}, \bibinfo {author} {\bibfnamefont {G.}~\bibnamefont {Coupier}},
  \bibinfo {author} {\bibfnamefont {A.}~\bibnamefont {Srivastav}}, \bibinfo
  {author} {\bibfnamefont {C.}~\bibnamefont {Minetti}}, \ and\ \bibinfo
  {author} {\bibfnamefont {T.}~\bibnamefont {Podgorski}},\ }\href@noop {}
  {\bibfield  {journal} {\bibinfo  {journal} {Phys. Rev. Lett.}\ }\textbf
  {\bibinfo {volume} {110}},\ \bibinfo {pages} {108101} (\bibinfo {year}
  {2013})}\BibitemShut {NoStop}%
\bibitem [{\citenamefont {Wei~Hou}\ \emph {et~al.}(2012)\citenamefont
  {Wei~Hou}, \citenamefont {Gan}, \citenamefont {Bhagat}, \citenamefont {Li},
  \citenamefont {Lim},\ and\ \citenamefont {Han}}]{WeiHou:2012is}%
  \BibitemOpen
  \bibfield  {author} {\bibinfo {author} {\bibfnamefont {H.}~\bibnamefont
  {Wei~Hou}}, \bibinfo {author} {\bibfnamefont {H.~Y.}\ \bibnamefont {Gan}},
  \bibinfo {author} {\bibfnamefont {A.~A.~S.}\ \bibnamefont {Bhagat}}, \bibinfo
  {author} {\bibfnamefont {L.~D.}\ \bibnamefont {Li}}, \bibinfo {author}
  {\bibfnamefont {C.~T.}\ \bibnamefont {Lim}}, \ and\ \bibinfo {author}
  {\bibfnamefont {J.}~\bibnamefont {Han}},\ }\href@noop {} {\bibfield
  {journal} {\bibinfo  {journal} {Biomicrofluidics}\ }\textbf {\bibinfo
  {volume} {6}},\ \bibinfo {pages} {024115} (\bibinfo {year}
  {2012})}\BibitemShut {NoStop}%
\bibitem [{\citenamefont {Namdee}\ \emph {et~al.}(2013)\citenamefont {Namdee},
  \citenamefont {Thompson}, \citenamefont {Charoenphol},\ and\ \citenamefont
  {Eniola-Adefeso}}]{Namdee:2013fc}%
  \BibitemOpen
  \bibfield  {author} {\bibinfo {author} {\bibfnamefont {K.}~\bibnamefont
  {Namdee}}, \bibinfo {author} {\bibfnamefont {A.~J.}\ \bibnamefont
  {Thompson}}, \bibinfo {author} {\bibfnamefont {P.}~\bibnamefont
  {Charoenphol}}, \ and\ \bibinfo {author} {\bibfnamefont {O.}~\bibnamefont
  {Eniola-Adefeso}},\ }\href@noop {} {\bibfield  {journal} {\bibinfo  {journal}
  {Langmuir}\ }\textbf {\bibinfo {volume} {29}},\ \bibinfo {pages} {2530}
  (\bibinfo {year} {2013})}\BibitemShut {NoStop}%
\bibitem [{\citenamefont {Thompson}\ \emph {et~al.}(2013)\citenamefont
  {Thompson}, \citenamefont {Mastria},\ and\ \citenamefont
  {Eniola-Adefeso}}]{Thompson:2013dm}%
  \BibitemOpen
  \bibfield  {author} {\bibinfo {author} {\bibfnamefont {A.~J.}\ \bibnamefont
  {Thompson}}, \bibinfo {author} {\bibfnamefont {E.~M.}\ \bibnamefont
  {Mastria}}, \ and\ \bibinfo {author} {\bibfnamefont {O.}~\bibnamefont
  {Eniola-Adefeso}},\ }\href@noop {} {\bibfield  {journal} {\bibinfo  {journal}
  {Biomaterials}\ }\textbf {\bibinfo {volume} {34}},\ \bibinfo {pages} {5863}
  (\bibinfo {year} {2013})}\BibitemShut {NoStop}%
\bibitem [{\citenamefont {Freund}(2007)}]{Freund:2007kx}%
  \BibitemOpen
  \bibfield  {author} {\bibinfo {author} {\bibfnamefont {J.~B.}\ \bibnamefont
  {Freund}},\ }\href@noop {} {\bibfield  {journal} {\bibinfo  {journal} {Phys.
  Fluids}\ }\textbf {\bibinfo {volume} {19}},\ \bibinfo {pages} {023301}
  (\bibinfo {year} {2007})}\BibitemShut {NoStop}%
\bibitem [{\citenamefont {Crowl}\ and\ \citenamefont
  {Fogelson}(2011)}]{Crowl:2011cf}%
  \BibitemOpen
  \bibfield  {author} {\bibinfo {author} {\bibfnamefont {L.}~\bibnamefont
  {Crowl}}\ and\ \bibinfo {author} {\bibfnamefont {A.~L.}\ \bibnamefont
  {Fogelson}},\ }\href@noop {} {\bibfield  {journal} {\bibinfo  {journal} {J.
  Fluid Mech.}\ }\textbf {\bibinfo {volume} {676}},\ \bibinfo {pages} {348}
  (\bibinfo {year} {2011})}\BibitemShut {NoStop}%
\bibitem [{\citenamefont {Zhao}\ and\ \citenamefont
  {Shaqfeh}(2011)}]{Zhao:2011do}%
  \BibitemOpen
  \bibfield  {author} {\bibinfo {author} {\bibfnamefont {H.}~\bibnamefont
  {Zhao}}\ and\ \bibinfo {author} {\bibfnamefont {E.~S.~G.}\ \bibnamefont
  {Shaqfeh}},\ }\href@noop {} {\bibfield  {journal} {\bibinfo  {journal} {Phys.
  Rev. E}\ }\textbf {\bibinfo {volume} {83}},\ \bibinfo {pages} {061924}
  (\bibinfo {year} {2011})}\BibitemShut {NoStop}%
\bibitem [{\citenamefont {Kumar}\ and\ \citenamefont
  {Graham}(2011)}]{Kumar:2011dd}%
  \BibitemOpen
  \bibfield  {author} {\bibinfo {author} {\bibfnamefont {A.}~\bibnamefont
  {Kumar}}\ and\ \bibinfo {author} {\bibfnamefont {M.~D.}\ \bibnamefont
  {Graham}},\ }\href@noop {} {\bibfield  {journal} {\bibinfo  {journal} {Phys.
  Rev. E}\ }\textbf {\bibinfo {volume} {84}},\ \bibinfo {pages} {066316}
  (\bibinfo {year} {2011})}\BibitemShut {NoStop}%
\bibitem [{\citenamefont {Fedosov}\ \emph {et~al.}(2012)\citenamefont
  {Fedosov}, \citenamefont {Fornleitner},\ and\ \citenamefont
  {Gompper}}]{Fedosov:2012dy}%
  \BibitemOpen
  \bibfield  {author} {\bibinfo {author} {\bibfnamefont {D.~A.}\ \bibnamefont
  {Fedosov}}, \bibinfo {author} {\bibfnamefont {J.}~\bibnamefont
  {Fornleitner}}, \ and\ \bibinfo {author} {\bibfnamefont {G.}~\bibnamefont
  {Gompper}},\ }\href@noop {} {\bibfield  {journal} {\bibinfo  {journal} {Phys.
  Rev. Lett.}\ }\textbf {\bibinfo {volume} {108}},\ \bibinfo {pages} {028104}
  (\bibinfo {year} {2012})}\BibitemShut {NoStop}%
\bibitem [{\citenamefont {Reasor}\ \emph {et~al.}(2013)\citenamefont {Reasor},
  \citenamefont {Mehrabadi}, \citenamefont {Ku},\ and\ \citenamefont
  {Aidun}}]{Reasor:2012ey}%
  \BibitemOpen
  \bibfield  {author} {\bibinfo {author} {\bibfnamefont {D.~A.}\ \bibnamefont
  {Reasor}}, \bibinfo {author} {\bibfnamefont {M.}~\bibnamefont {Mehrabadi}},
  \bibinfo {author} {\bibfnamefont {D.~N.}\ \bibnamefont {Ku}}, \ and\ \bibinfo
  {author} {\bibfnamefont {C.~K.}\ \bibnamefont {Aidun}},\ }\href@noop {}
  {\bibfield  {journal} {\bibinfo  {journal} {Ann. Biomed. Eng.}\ }\textbf
  {\bibinfo {volume} {41}},\ \bibinfo {pages} {238} (\bibinfo {year}
  {2013})}\BibitemShut {NoStop}%
\bibitem [{\citenamefont {Zhao}\ \emph {et~al.}(2012)\citenamefont {Zhao},
  \citenamefont {Shaqfeh},\ and\ \citenamefont {Narsimhan}}]{Zhao:2012gg}%
  \BibitemOpen
  \bibfield  {author} {\bibinfo {author} {\bibfnamefont {H.}~\bibnamefont
  {Zhao}}, \bibinfo {author} {\bibfnamefont {E.~S.~G.}\ \bibnamefont
  {Shaqfeh}}, \ and\ \bibinfo {author} {\bibfnamefont {V.}~\bibnamefont
  {Narsimhan}},\ }\href@noop {} {\bibfield  {journal} {\bibinfo  {journal}
  {Phys. Fluids}\ }\textbf {\bibinfo {volume} {24}},\ \bibinfo {pages} {011902}
  (\bibinfo {year} {2012})}\BibitemShut {NoStop}%
\bibitem [{\citenamefont {Fedosov}\ \emph {et~al.}(2013)\citenamefont
  {Fedosov}, \citenamefont {Dao}, \citenamefont {Karniadakis},\ and\
  \citenamefont {Suresh}}]{Fedosov:2013ie}%
  \BibitemOpen
  \bibfield  {author} {\bibinfo {author} {\bibfnamefont {D.~A.}\ \bibnamefont
  {Fedosov}}, \bibinfo {author} {\bibfnamefont {M.}~\bibnamefont {Dao}},
  \bibinfo {author} {\bibfnamefont {G.~E.}\ \bibnamefont {Karniadakis}}, \ and\
  \bibinfo {author} {\bibfnamefont {S.}~\bibnamefont {Suresh}},\ }\href@noop {}
  {\bibfield  {journal} {\bibinfo  {journal} {Ann. Biomed. Eng.}\ }\textbf
  {\bibinfo {volume} {42}},\ \bibinfo {pages} {368} (\bibinfo {year}
  {2013})}\BibitemShut {NoStop}%
\bibitem [{\citenamefont {Vahidkhah}\ \emph {et~al.}(2014)\citenamefont
  {Vahidkhah}, \citenamefont {Diamond},\ and\ \citenamefont
  {Bagchi}}]{Vahidkhah:2014hy}%
  \BibitemOpen
  \bibfield  {author} {\bibinfo {author} {\bibfnamefont {K.}~\bibnamefont
  {Vahidkhah}}, \bibinfo {author} {\bibfnamefont {S.~L.}\ \bibnamefont
  {Diamond}}, \ and\ \bibinfo {author} {\bibfnamefont {P.}~\bibnamefont
  {Bagchi}},\ }\href@noop {} {\bibfield  {journal} {\bibinfo  {journal}
  {Biophys. J.}\ }\textbf {\bibinfo {volume} {106}},\ \bibinfo {pages} {2529}
  (\bibinfo {year} {2014})}\BibitemShut {NoStop}%
\bibitem [{\citenamefont {Kumar}\ \emph {et~al.}(2014)\citenamefont {Kumar},
  \citenamefont {Henriquez~Rivera},\ and\ \citenamefont
  {Graham}}]{Kumar:2013tu}%
  \BibitemOpen
  \bibfield  {author} {\bibinfo {author} {\bibfnamefont {A.}~\bibnamefont
  {Kumar}}, \bibinfo {author} {\bibfnamefont {R.}~\bibnamefont
  {Henriquez~Rivera}}, \ and\ \bibinfo {author} {\bibfnamefont {M.~D.}\
  \bibnamefont {Graham}},\ }\href@noop {} {\bibfield  {journal} {\bibinfo
  {journal} {J. Fluid Mech.}\ }\textbf {\bibinfo {volume} {738}},\ \bibinfo
  {pages} {423} (\bibinfo {year} {2014})}\BibitemShut {NoStop}%
\bibitem [{\citenamefont {Da~Cunha}\ and\ \citenamefont
  {Hinch}(1996)}]{cunha96}%
  \BibitemOpen
  \bibfield  {author} {\bibinfo {author} {\bibfnamefont {F.~R.}\ \bibnamefont
  {Da~Cunha}}\ and\ \bibinfo {author} {\bibfnamefont {E.~J.}\ \bibnamefont
  {Hinch}},\ }\href@noop {} {\bibfield  {journal} {\bibinfo  {journal} {J.
  Fluid Mech.}\ }\textbf {\bibinfo {volume} {309}},\ \bibinfo {pages} {211}
  (\bibinfo {year} {1996})}\BibitemShut {NoStop}%
\bibitem [{\citenamefont {Li}\ and\ \citenamefont
  {Pozrikidis}(2000)}]{Li:2000wv}%
  \BibitemOpen
  \bibfield  {author} {\bibinfo {author} {\bibfnamefont {X.~F.}\ \bibnamefont
  {Li}}\ and\ \bibinfo {author} {\bibfnamefont {C.}~\bibnamefont
  {Pozrikidis}},\ }\href@noop {} {\bibfield  {journal} {\bibinfo  {journal}
  {Int. J. Multiphase Flow}\ }\textbf {\bibinfo {volume} {26}},\ \bibinfo
  {pages} {1247} (\bibinfo {year} {2000})}\BibitemShut {NoStop}%
\bibitem [{\citenamefont {Zurita-Gotor}\ \emph {et~al.}(2012)\citenamefont
  {Zurita-Gotor}, \citenamefont {Blawzdziewicz},\ and\ \citenamefont
  {Wajnryb}}]{zurita12}%
  \BibitemOpen
  \bibfield  {author} {\bibinfo {author} {\bibfnamefont {M.}~\bibnamefont
  {Zurita-Gotor}}, \bibinfo {author} {\bibfnamefont {J.}~\bibnamefont
  {Blawzdziewicz}}, \ and\ \bibinfo {author} {\bibfnamefont {E.}~\bibnamefont
  {Wajnryb}},\ }\href@noop {} {\bibfield  {journal} {\bibinfo  {journal} {Phys.
  Rev. Lett.}\ }\textbf {\bibinfo {volume} {108}},\ \bibinfo {pages} {068301}
  (\bibinfo {year} {2012})}\BibitemShut {NoStop}%
\bibitem [{\citenamefont {Smart}\ and\ \citenamefont
  {Leighton}(1991)}]{Smart:1991vp}%
  \BibitemOpen
  \bibfield  {author} {\bibinfo {author} {\bibfnamefont {J.~R.}\ \bibnamefont
  {Smart}}\ and\ \bibinfo {author} {\bibfnamefont {D.~T.}\ \bibnamefont
  {Leighton}},\ }\href@noop {} {\bibfield  {journal} {\bibinfo  {journal}
  {Phys. Fluids A}\ }\textbf {\bibinfo {volume} {3}},\ \bibinfo {pages} {21}
  (\bibinfo {year} {1991})}\BibitemShut {NoStop}%
\bibitem [{\citenamefont {Ma}\ and\ \citenamefont {Graham}(2005)}]{Ma:2005dw}%
  \BibitemOpen
  \bibfield  {author} {\bibinfo {author} {\bibfnamefont {H.}~\bibnamefont
  {Ma}}\ and\ \bibinfo {author} {\bibfnamefont {M.~D.}\ \bibnamefont
  {Graham}},\ }\href@noop {} {\bibfield  {journal} {\bibinfo  {journal} {Phys.
  Fluids}\ }\textbf {\bibinfo {volume} {17}},\ \bibinfo {pages} {083103}
  (\bibinfo {year} {2005})}\BibitemShut {NoStop}%
\bibitem [{\citenamefont {Kumar}\ and\ \citenamefont
  {Graham}(2012{\natexlab{b}})}]{Kumar:2012ie}%
  \BibitemOpen
  \bibfield  {author} {\bibinfo {author} {\bibfnamefont {A.}~\bibnamefont
  {Kumar}}\ and\ \bibinfo {author} {\bibfnamefont {M.~D.}\ \bibnamefont
  {Graham}},\ }\href@noop {} {\bibfield  {journal} {\bibinfo  {journal} {Phys.
  Rev. Lett.}\ }\textbf {\bibinfo {volume} {109}},\ \bibinfo {pages} {108102}
  (\bibinfo {year} {2012}{\natexlab{b}})}\BibitemShut {NoStop}%
\bibitem [{\citenamefont {Narsimhan}\ \emph {et~al.}(2013)\citenamefont
  {Narsimhan}, \citenamefont {Zhao},\ and\ \citenamefont
  {Shaqfeh}}]{Narsimhan:2013jk}%
  \BibitemOpen
  \bibfield  {author} {\bibinfo {author} {\bibfnamefont {V.}~\bibnamefont
  {Narsimhan}}, \bibinfo {author} {\bibfnamefont {H.}~\bibnamefont {Zhao}}, \
  and\ \bibinfo {author} {\bibfnamefont {E.~S.~G.}\ \bibnamefont {Shaqfeh}},\
  }\href@noop {} {\bibfield  {journal} {\bibinfo  {journal} {Phys. Fluids}\
  }\textbf {\bibinfo {volume} {25}},\ \bibinfo {pages} {061901} (\bibinfo
  {year} {2013})}\BibitemShut {NoStop}%
\bibitem [{\citenamefont {Boal}(2012)}]{Boal:2012wq}%
  \BibitemOpen
  \bibfield  {author} {\bibinfo {author} {\bibfnamefont {D.}~\bibnamefont
  {Boal}},\ }\href@noop {} {\emph {\bibinfo {title} {{Mechanics of the
  Cell}}}},\ \bibinfo {edition} {2nd}\ ed.\ (\bibinfo  {publisher} {Cambridge
  University Press},\ \bibinfo {address} {Cambridge},\ \bibinfo {year}
  {2012})\BibitemShut {NoStop}%
\bibitem [{\citenamefont {Pranay}\ \emph {et~al.}(2012)\citenamefont {Pranay},
  \citenamefont {Henriquez~Rivera},\ and\ \citenamefont
  {Graham}}]{Pranay:2012ku}%
  \BibitemOpen
  \bibfield  {author} {\bibinfo {author} {\bibfnamefont {P.}~\bibnamefont
  {Pranay}}, \bibinfo {author} {\bibfnamefont {R.~G.}\ \bibnamefont
  {Henriquez~Rivera}}, \ and\ \bibinfo {author} {\bibfnamefont {M.~D.}\
  \bibnamefont {Graham}},\ }\href@noop {} {\bibfield  {journal} {\bibinfo
  {journal} {Phys. Fluids}\ }\textbf {\bibinfo {volume} {24}},\ \bibinfo
  {pages} {061902} (\bibinfo {year} {2012})}\BibitemShut {NoStop}%
\bibitem [{\citenamefont {Phillips}\ \emph {et~al.}(1992)\citenamefont
  {Phillips}, \citenamefont {Armstrong}, \citenamefont {Brown}, \citenamefont
  {Graham},\ and\ \citenamefont {Abbott}}]{PHILLIPS:1992vy}%
  \BibitemOpen
  \bibfield  {author} {\bibinfo {author} {\bibfnamefont {R.~J.}\ \bibnamefont
  {Phillips}}, \bibinfo {author} {\bibfnamefont {R.~C.}\ \bibnamefont
  {Armstrong}}, \bibinfo {author} {\bibfnamefont {R.~A.}\ \bibnamefont
  {Brown}}, \bibinfo {author} {\bibfnamefont {A.~L.}\ \bibnamefont {Graham}}, \
  and\ \bibinfo {author} {\bibfnamefont {J.~R.}\ \bibnamefont {Abbott}},\
  }\href@noop {} {\bibfield  {journal} {\bibinfo  {journal} {Phys. Fluids A}\
  }\textbf {\bibinfo {volume} {4}},\ \bibinfo {pages} {30} (\bibinfo {year}
  {1992})}\BibitemShut {NoStop}%
\bibitem [{\citenamefont {Leighton}\ and\ \citenamefont
  {Acrivos}(1987)}]{LEIGHTON:1987uo}%
  \BibitemOpen
  \bibfield  {author} {\bibinfo {author} {\bibfnamefont {D.~T.}\ \bibnamefont
  {Leighton}}\ and\ \bibinfo {author} {\bibfnamefont {A.}~\bibnamefont
  {Acrivos}},\ }\href@noop {} {\bibfield  {journal} {\bibinfo  {journal} {J.
  Fluid Mech.}\ }\textbf {\bibinfo {volume} {181}},\ \bibinfo {pages} {415}
  (\bibinfo {year} {1987})}\BibitemShut {NoStop}%
\bibitem [{\citenamefont {Hudson}(2003)}]{Hudson:2003df}%
  \BibitemOpen
  \bibfield  {author} {\bibinfo {author} {\bibfnamefont {S.~D.}\ \bibnamefont
  {Hudson}},\ }\href@noop {} {\bibfield  {journal} {\bibinfo  {journal} {Phys.
  Fluids}\ }\textbf {\bibinfo {volume} {15}},\ \bibinfo {pages} {1106}
  (\bibinfo {year} {2003})}\BibitemShut {NoStop}%
\bibitem [{\citenamefont {Kumar}\ and\ \citenamefont
  {Graham}(2012{\natexlab{c}})}]{Kumar:2012ev}%
  \BibitemOpen
  \bibfield  {author} {\bibinfo {author} {\bibfnamefont {A.}~\bibnamefont
  {Kumar}}\ and\ \bibinfo {author} {\bibfnamefont {M.~D.}\ \bibnamefont
  {Graham}},\ }\href@noop {} {\bibfield  {journal} {\bibinfo  {journal} {J.
  Comput. Phys.}\ }\textbf {\bibinfo {volume} {231}},\ \bibinfo {pages} {6682}
  (\bibinfo {year} {2012}{\natexlab{c}})}\BibitemShut {NoStop}%
\bibitem [{\citenamefont {Schmid-Sch{\"o}nbein}\ \emph
  {et~al.}(1980)\citenamefont {Schmid-Sch{\"o}nbein}, \citenamefont {Shih},\
  and\ \citenamefont {Chien}}]{schmid1980morphometry}%
  \BibitemOpen
  \bibfield  {author} {\bibinfo {author} {\bibfnamefont {G.~W.}\ \bibnamefont
  {Schmid-Sch{\"o}nbein}}, \bibinfo {author} {\bibfnamefont {Y.~Y.}\
  \bibnamefont {Shih}}, \ and\ \bibinfo {author} {\bibfnamefont
  {S.}~\bibnamefont {Chien}},\ }\href@noop {} {\bibfield  {journal} {\bibinfo
  {journal} {Blood}\ }\textbf {\bibinfo {volume} {56}},\ \bibinfo {pages} {866}
  (\bibinfo {year} {1980})}\BibitemShut {NoStop}%
\bibitem [{\citenamefont {Paulus}(1975)}]{paulus1975platelet}%
  \BibitemOpen
  \bibfield  {author} {\bibinfo {author} {\bibfnamefont {J.~M.}\ \bibnamefont
  {Paulus}},\ }\href@noop {} {\bibfield  {journal} {\bibinfo  {journal}
  {Blood}\ }\textbf {\bibinfo {volume} {46}},\ \bibinfo {pages} {321} (\bibinfo
  {year} {1975})}\BibitemShut {NoStop}%
\bibitem [{\citenamefont {Schmid-Sch{\"o}nbein}\ \emph
  {et~al.}(1981)\citenamefont {Schmid-Sch{\"o}nbein}, \citenamefont {Sung},
  \citenamefont {T{\"o}zeren}, \citenamefont {Skalak},\ and\ \citenamefont
  {Chien}}]{schmid1981passive}%
  \BibitemOpen
  \bibfield  {author} {\bibinfo {author} {\bibfnamefont {G.~W.}\ \bibnamefont
  {Schmid-Sch{\"o}nbein}}, \bibinfo {author} {\bibfnamefont {K.~L.}\
  \bibnamefont {Sung}}, \bibinfo {author} {\bibfnamefont {H.}~\bibnamefont
  {T{\"o}zeren}}, \bibinfo {author} {\bibfnamefont {R.}~\bibnamefont {Skalak}},
  \ and\ \bibinfo {author} {\bibfnamefont {S.}~\bibnamefont {Chien}},\
  }\href@noop {} {\bibfield  {journal} {\bibinfo  {journal} {Biophys. J.}\
  }\textbf {\bibinfo {volume} {36}},\ \bibinfo {pages} {243} (\bibinfo {year}
  {1981})}\BibitemShut {NoStop}%
\bibitem [{\citenamefont {Lam}\ \emph {et~al.}(2011)\citenamefont {Lam},
  \citenamefont {Chaudhuri}, \citenamefont {Crow}, \citenamefont {Webster},
  \citenamefont {Kita}, \citenamefont {Huang}, \citenamefont {Fletcher} \emph
  {et~al.}}]{lam2011mechanics}%
  \BibitemOpen
  \bibfield  {author} {\bibinfo {author} {\bibfnamefont {W.~A.}\ \bibnamefont
  {Lam}}, \bibinfo {author} {\bibfnamefont {O.}~\bibnamefont {Chaudhuri}},
  \bibinfo {author} {\bibfnamefont {A.}~\bibnamefont {Crow}}, \bibinfo {author}
  {\bibfnamefont {K.~D.}\ \bibnamefont {Webster}}, \bibinfo{author} {\bibnamefont {T.}~\bibnamefont {Li}}, \bibinfo {author}
  {\bibfnamefont {A.}~\bibnamefont {Kita}}, \bibinfo {author} {\bibfnamefont
  {J.}~\bibnamefont {Huang}},\ and\ \bibinfo {author} {\bibfnamefont {D.~A.}\
  \bibnamefont {Fletcher}},\ }\href@noop {} {\bibfield
  {journal} {\bibinfo  {journal} {Nat. Mater.}\ }\textbf {\bibinfo {volume}
  {10}},\ \bibinfo {pages} {61} (\bibinfo {year} {2011})}\BibitemShut {NoStop}%
\end{thebibliography}
\end{document}